\documentstyle [12pt]{article}
\makeatletter \oddsidemargin  0in \evensidemargin 0in
\textwidth16cm \RequirePackage[dvips]{graphicx} \textheight 20.5cm
\newcommand{\singlespacing}{\let\CS=\@currsize\renewcommand{\baselinestretch}{1.5}\tiny\CS}
\makeatletter \oddsidemargin  -.1in \evensidemargin -.1in
\newcommand{\doublespacing}{\let\CS=\@currsize\renewcommand{\baselinestretch}{1.35}\tiny\CS}

\def\@citex[#1]#2{\if@filesw\immediate\write\@auxout{\string\citation{#2}}\fi
  \def\@citea{}\@cite{\@for\@citeb:=#2\do
    {\@citea\def\@citea{,\linebreak[0]\hskip0pt plus .2em}%
      \@ifundefined{b@\@citeb}%
    {{\bf ?}\@warning{Citation `\@citeb' on page \thepage\space undefined}}%
      \hbox{\csname b@\@citeb\endcsname}}}{#1}}

\newtheorem{rule-def}[theorem]{Rule}
\begin{document}
\title{\bf Quantum Cloning, Bell's Inequality and Teleportation }
\author{S.Adhikari $^{1}$, N.Ganguly $^{2,3}$\thanks{Corresponding author:
nirmanganguly@rediffmail.com }, I.Chakrabarty $^{2,3}$,B.S.Choudhury $^{3}$\\
$^1$ S.N.Bose National Centre for Basic Sciences, Salt lake, WestBengal, India.\\
$^2$ Heritage Institute of Technology, Kolkata-107, West Bengal, India\\
$^3$ Bengal Engineering and Science University, Howrah, West
Bengal, India }
\date{}
\maketitle{}
\begin{abstract}
We analyze a possibility of using the two qubit output state from
Buzek-Hillery quantum copying machine (not necessarily universal
quantum cloning machine) as a teleportation channel. We show that
there is a range of values of the machine parameter $\xi$ for
which the two qubit output state is entangled and violates
Bell-CHSH inequality and for a different range it remains
entangled but does not violate Bell-CHSH inequality. Further we
observe that for certain values of the machine parameter the
two-qubit mixed state can be used as a teleportation channel. The
use of the output state from the Buzek-Hillery cloning machine as
a teleportation channel provides an additional appeal to the
cloning machine and motivation of our present work.\\
\end{abstract}

\section{Introduction } In 1993 an international team of six
scientists including Charles Bennett \cite{ben} discovered a new
aspect of quantum inseparability - teleportation. Teleportation is
purely based on classical information and non-classical
Einstein-Podolsky-Rosen(EPR) correlations. The basic idea is to use
a pair of particles in a singlet state shared by distant partners Alice and
Bob to perfom successful teleportation of an arbitrary qubit
from the sender Alice to the receiver Bob. There was a question what value of fidelity of
transmission of an unknown state can ensure us about non-classical
character of the state forming the quantum channel. It has been
shown that the purely classical channel can
give at most $\textit{F}=\frac{2}{3}$ \cite{ng,pop,mp}.\\
Then Popescu raised basic questions concerning a possible
relation between teleportation, Bell-CHSH inequalities \cite{chsh} and
inseparability: "What is the exact relation between Bell's
inequalities violation and teleportation?"\cite{pop}. In this paper, we probe these relations for the
two qubit mixed entangled state arising from the B-H quantum cloning machine.\\
Bell's inequalities are relations between conditional
probabilities valid under the locality assumption. Hence, $\textit{a
 priori}$ they have nothing to do with quantum mechanics. However,
it is the fact that quantum mechanics predicts a violation of
 these conditions that makes them interesting.  Werner \cite{werner}
gave an example of an entangled state described by the density
operator $\rho_{W}=p|\psi^{-}\rangle\langle
\psi^{-}|+\frac{1-p}{4}I$, where
$|\psi^{-}\rangle=\frac{1}{\sqrt{2}}(|01\rangle-|10\rangle)$ and I
is the identity operator in the 4-dimensional Hilbert space, which
does not violate the Bell's inequality for
$\frac{1}{3}<p<\frac{1}{\sqrt2}$. Interestingly in this work, we
also find an example of an entangled state that does not violate
the Bell's inequality. A natural question arises in concern with
teleportation whether states which violate
 Bell-CHSH inequalities are suitable for teleportation. Horodecki $\textit{et
 al}$ \cite{h3} showed that any mixed two spin-$\frac{1}{2}$ state which
 violates the Bell-CHSH inequalities is suitable for teleportation. \\
 With the advent of quantum cloning machine we were introduced to
 an entanglement between the output and input states. The proposition of
 universal quantum copying machine by Buzek and Hillery \cite{bh}
 presented us a cloning machine which is independent of the input
 state. However, as was quite expected the copy and original which
 appear at the output remained entangled.\\
 Our investigation starts from the idea of using the two qubit entangled state which comes as a output of the
 B-H cloning machine as a teleportation channel. Central to this idea is the possible utility of a mixed
 entangled state as a teleportation channel, which motivates us for this work. Our analysis results in a certain range
 of the machine parameter $\xi$ for which the entangled state can be used as a teleportation channel.
 For $\xi=\frac{1}{2}$, the output state reduces to a maximally entangled pure state which can be used as a
 teleportation channel faithfully as expected. Altogether this work provides an additional
 appeal for the B-H cloning machine.\\
 Our work is organised as follows: In the following section, we discuss the
 viability of using the output state given by the B-H cloning
 machine as a teleportation channel. Lastly we summarize our work.
\section{Analysis of the output of Buzek-Hillery copying machine}
In this section we study the viability of the entangled output
copies of Buzek-Hillery cloning machine  \cite{bh} as a teleportation channel. \\
The cloning transformation for copying procedure \cite{bh} is
given by
\begin{eqnarray}
|0\rangle_{a}|0\rangle_{b}|Q\rangle_{x}\longrightarrow|0\rangle_{a}|0\rangle_{b}|Q_0\rangle_{x}
+[|0\rangle_a|1\rangle_b+|1\rangle_a|0\rangle_b]|Y_0\rangle_x\\
|1\rangle_{a}|0\rangle_{b}|Q\rangle_{x}\longrightarrow|1\rangle_{a}|1\rangle_{b}|Q_1\rangle_{x}
+[|0\rangle_a|1\rangle_b+|1\rangle_a|0\rangle_b]|Y_1\rangle_x
\end{eqnarray}
The unitarity and the orthogonality of the cloning transformation
gives
\begin{eqnarray}
_{x}\langle Q_i|Q_i\rangle_{x}+2_{x}\langle Y_i|Y_i\rangle_{x}
=1~~~~(i=0,1)\\
 _{x}\langle Y_0|Y_1\rangle_{x}=_{x}\langle
Y_1|Y_0\rangle_{x}=0
\end{eqnarray}
Here the copying machine state vectors $|Y_i\rangle_{x}$ and
$|Q_i\rangle_{x}$ are assumed to be mutually orthogonal, so are
the state vectors $\{|Q_0\rangle,|Q_1\rangle \}$.\\
Let us consider a quantum state which is to be cloned
\begin{eqnarray}
|\chi\rangle= \alpha|0\rangle+\beta|1\rangle
\end{eqnarray}
where $\alpha^{2}+\beta^{2}=1$.\\
Here we confine ourselves to a limited class of input states (5),
where $\alpha$ and $\beta$ are real.\\
After using the cloning transformation (1-2) on quantum state (5)
and tracing out the machine state vector, the two qubit reduced
density operator describing the two clones is given by
\begin{eqnarray}
\rho_{ab}^{out}=\alpha^2(1-2\xi)|00\rangle\langle
00|+\frac{\alpha\beta}{\sqrt{2}}(1-2\xi)|00\rangle\langle
+|+\frac{\alpha\beta}{\sqrt{2}}(1-2\xi)|+\rangle\langle
00|\nonumber\\+2\xi|+\rangle\langle
+|+\frac{\alpha\beta}{\sqrt{2}}(1-2\xi)|+\rangle\langle 11|
+\frac{\alpha\beta}{\sqrt{2}}(1-2\xi)|11\rangle\langle
+|+\beta^2(1-2\xi)|11\rangle\langle 11|
\end{eqnarray}
where we have used the following notations\\
$_{x}\langle Y_0|Y_0\rangle_{x}$=$_{x}\langle Y_1|Y_1\rangle_{x}=
\xi$, $_{x}\langle Y_0 |Q_1\rangle_{x}$=$_{x}\langle Q_0
|Y_1\rangle_{x}$=$_{x}\langle Q_1|Y_0\rangle_{x}$=$_{x}\langle Y_1
|Q_0\rangle_{x}$=$ \
\frac{\eta}{2}$,\\
$|+\rangle=\frac{1}{\sqrt{2}}(|01\rangle+|10\rangle)$.\\
Also we have used the relation $\eta=1-2\xi$ in equation (6) to
make the distortion between the state $\rho^{id}$ and the one
qubit reduced
state $\rho_{a}^{out}$ input state independent.\\
The necessary and sufficient condition for a state $\rho$  to be
inseparable is that at least one of the eigen values of the
partially transposed operator defined as
$\rho^{T}_{m\mu,n\nu}=\rho_{m\nu,n\mu}$ is negative
\cite{peres,3h}.
This is equivalent to the condition that at least one of the two determinants\\
$W_{3}= \begin{array}{|ccc|}
  \rho_{00,00} & \rho_{01,00} & \rho_{00,10} \\
  \rho_{00,01} & \rho_{01,01} & \rho_{00,11} \\
  \rho_{10,00} & \rho_{11,00} & \rho_{10,10}
\end{array}$ and $W_{4}=\begin{array}{|cccc|}
   \rho_{00,00} & \rho_{01,00} & \rho_{00,10} & \rho_{01,10}\\
  \rho_{00,01} & \rho_{01,01} & \rho_{00,11} & \rho_{01,11} \\
  \rho_{10,00} & \rho_{11,00} & \rho_{10,10} & \rho_{11,10} \\
  \rho_{10,01} & \rho_{11,01} & \rho_{10,11} & \rho_{11,11}
\end{array}$\\
is negative.\\
Now we investigate the inseparability of two qubit density
operator $\rho^{out}_{ab}$ for different intervals of the machine parameter $\xi$.\\
For the density matrix $\rho^{out}_{ab}$, we calculate the
determinants $W_{3}$ and $W_{4}$, which are given by
\begin{eqnarray}
W_{3}= \frac{\alpha^{2}\xi(1-2\xi)}{2}[2\xi-\beta^{2}(1-2\xi)],~
W_{4}=\frac{1}{2}[\alpha^{2}\beta^{2}\xi(1-2\xi)^{2}(6\xi-1)-2\xi^{4}]
\end{eqnarray}
Further we note that the relation $\eta=1-2\xi$ reduces the Schwarz inequality $\eta\leq2(\xi-2\xi^{2})^{\frac{1}{2}}$
to the inequality $\frac{1}{6}\leq\xi\leq\frac{1}{2}$. So we consider the following cases :\\
(i) For $\xi=\frac{1}{6}$ and $\xi=\frac{1}{2}$, it is clear that
$W_{4}<0$ for all $\alpha,~\beta$ and hence
the two qubit density operator $\rho^{out}_{ab}$ is inseparable for all $\alpha,~\beta$.\\
(ii)In the interval $\frac{1}{6}<\xi<\frac{1}{2}$,there exist sub
intervals of the machine parameter $\xi$ for which the density
operator $\rho^{out}_{ab}$ is inseparable for some
interval of $\alpha^{2}$. Now we discuss two subcases below where the range of $\alpha^{2}$ is given for which $\rho^{out}_{ab}$ is inseparable.\\
(iia) In the interval $0<\alpha^{2}<\frac{1-4\xi}{1-2\xi}$, the
determinant $W_{3}<0$ and hence $\rho^{out}_{ab}$ is inseparable
when the machine parameter lies in the interval $\frac{1}{6}<\xi<\frac{1}{4}$.\\
(iib) Moreover in the interval
$\frac{1-4\xi}{1-2\xi}<\alpha^{2}<\frac{1}{2}-\frac{\sqrt{A^{2}-8A\xi^{4}}}{2A}$
where $A=\xi(6\xi-1)(1-2\xi)^{2}$,  we find that the determinant
$W_{3}>0$ but $W_{4}<0$ and hence in this case the state
$\rho^{out}_{ab}$ is inseparable when the machine parameter $\xi$ lies in the range $\frac{3-\sqrt{5}}{4}<\xi<\frac{1}{4}$.\\
An arbitrary state of a two qubit system can be represented as
\begin{eqnarray}
\rho=\frac{1}{4}[I\otimes I+r.\sigma \otimes I+I\otimes s.
\sigma+\sum_{i,j=1}^3 \lambda_{ij}~\sigma_i\otimes\sigma_j]
\end{eqnarray}
where $\rho$ acts on the Hilbert space $\emph{H}= \emph{H}_1
\otimes \emph{H}_2$, $\textit{I}$ stands for identity operator,
$\{\sigma_i\}_{i=1}^3$ are the standard Pauli matrices, $r,s$ are
vectors in $R^3$, $r.\sigma=\sum_{i=1}^{3}r_i\sigma_i$ and
$s.\sigma=\sum_{i=1}^{3}s_i\sigma_i$. The coefficients
$\lambda_{ij}=Tr(\rho~\sigma_i\otimes \sigma_j)$ form a real
$3\times3$ matrix which we shall denote by $C(\rho)$.\\
Further, it is known that the state which does not violate Bell-CHSH
inequality must satisfy $M(\rho)\leq1$, where
$M(\rho)=max_{i>j}(u_{i}+u_{j})$, $u_{i}$ and $u_{j}$ are the eigen values of $U=C^{t}(\rho)C(\rho)$ \cite{3h1}\\
In our case, the elements of the correlation matrix
$C(\rho_{ab}^{out})$ obtained for the density operator $\rho_{ab}^{out}$ are
\begin{eqnarray}
c_{11}=2\xi, c_{12}=0, c_{13}=0, c_{21}=0, c_{22}=-2\xi, c_{23}=0,
c_{31}=0, c_{32}=0, c_{33}=1-4\xi
\end{eqnarray}
The eigen values of the matrix $U=C^{t}(\rho_{ab}^{out})C(\rho_{ab}^{out})$ are\\
$u_1=u_2=4\xi^2, u_3=1-8\xi+16\xi^2$.\\
Now we are in a position to discuss the different cases which
establishes a relation between Bell-CHSH inequality violation and
inseparability. \\
\textbf{Case-I:} If $u_{1}>u_{3}$ then the machine parameter
$\xi$ lies between $\frac{1}{6}$ and $\frac{1}{2}$.\\
In $\frac{1}{6}<\xi<\frac{1}{2\sqrt{2}}, M(\rho)=8\xi^2<1$ and
hence in this interval the two qubit density operator doesn't
violate the Bell-CHSH inequality. However in the interval
$\frac{1}{2\sqrt{2}}<\xi<\frac{1}{2}, M(\rho)=8\xi^2>1$ and hence
it violates the Bell-CHSH inequality\\
\textbf{Case-II:} If $u_{1}=u_{2}=u_{3}=\frac{1}{9}$ (which
happens when $\xi=\frac{1}{6}$) , then $M(\rho)<1$ and hence the
two qubit
entangled state does not violate the Bell-CHSH inequality.\\
\textbf{Case-III}: If $u_{1}=u_{2}=u_{3}=1$ (which happens when
$\xi=\frac{1}{2}$), then $M(\rho)>1$, and hence the two qubit
entangled state violates the Bell-CHSH inequality.\\For any mixed
spin $\frac{1}{2}$ state $\rho$ the teleportation fidelity
\cite{h3} amounts to $\texttt{F}^{max}= \frac{1}{2}(1+\frac{1}{3}
N(\rho))$, where $N(\rho)= \sum_{i=1}^{3}\sqrt{u_{i}}$.
\\We split the interval $\frac{1}{6}\leq\xi<\frac{1}{2}$ into two sub-intervals $\frac{1}{6}\leq\xi\leq\frac{1}{4}$ and
$\frac{1}{4}<\xi<\frac{1}{2}$. We notice that in the interval
$\frac{1}{6}\leq\xi\leq\frac{1}{4}$ , the teleportation fidelity
does not cross the classical limit $\frac{2}{3}$. Hence for this
interval we cannot use $\rho_{ab}^{out}$ as a teleportation
channel. However, in the interval $\frac{1}{4}<\xi<\frac{1}{2}$
the fidelity $\texttt{F}^{max}$ crosses $\frac{2}{3}$. So for this interval of $\xi$, the output state $\rho_{ab}^{out}$
can be used as a teleportation channel.\\
We note that in the case $\xi=\frac{1}{2}$, $\rho_{ab}^{out}$
reduces to a maximally pure entangled state which can be used in
teleportation with fidelity 1.
\section{Conclusions }
To summarize, we have cited an example of a two qubit entangled
state (output of the Buzek-Hillery cloning machine) which does not
violate Bell-CHSH inequality for some interval of the machine
parameter $\xi$. Also we have shown that the two qubit entangled
state does violate Bell-CHSH inequality for some interval of the
machine parameter $\xi$ different from the previous one. In some
of the cases, we found that the two qubit entangled state does act
as a useful quantum channel for teleportation protocol. This work
adds a special feature to the Buzek-Hillery quantum cloning
machine.

\end{document}